\documentclass{ws-mpla}   

\usepackage{url}

\usepackage{hyperref}
\usepackage{overcite}

\usepackage{graphicx}
\usepackage{amsmath}
\usepackage{slashed}
\usepackage[usenames]{color}
\usepackage{longtable}

\usepackage{tabularx}
\usepackage{booktabs}
\usepackage{siunitx}

\def\beq{\begin{equation}}
\def\eeq{\end{equation}}
\def\bea{\begin{eqnarray}}
\def\eea{\end{eqnarray}}
\def\D0{D\O }

\begin{document}
\newcommand{\mx} {\ensuremath{m_{X}}\xspace}
\newcommand{\mxqsq} {\ensuremath{(m_{X}, q^2)}\xspace}
\newcommand {\pplus}  {\ensuremath{P_{+}}\xspace}
\newcommand{\elsmax}{\ensuremath{(E_{\ell},s_{\mathrm{h}}^{\mathrm{max}})}}
\newcommand{\smax}{\ensuremath{s_{\mathrm{h}}^{\mathrm{max}}}}
\newcommand{\el} {\ensuremath{E_{\ell}}\xspace}
\newcommand {\mb}{\ensuremath{m_b}}
\newcommand {\mc}{\ensuremath{m_c}\xspace}
\newcommand{\btou}{\ensuremath{\bar{B} \to X_u\, l\,  \bar{\nu}_l}}
\newcommand{\btoc}{\ensuremath{\bar{B}\to X_c\, l\,  \bar{\nu}_l}}
\newcommand{\bsg}{\ensuremath{\bar{B}\to X_s\, \gamma}}
\newcommand{\Vcb} {\ensuremath{|V_{cb}|}}
\newcommand{\Vub} {\ensuremath{|V_{ub}| }}
\newcommand{\Vtb} {\ensuremath{|V_{tb}| }}

\newcommand*\xbar[1]{%
  \hbox{%
    \vbox{%
      \hrule height 0.5pt 
      \kern0.5ex
      \hbox{%
        \kern-0.1em
        \ensuremath{#1}%
        \kern-0.1em
      }%
    }%
  }%
}

\def\st{\scriptstyle}
\def\sst{\scriptscriptstyle}
\def\mco{\multicolumn}
\def\epp{\epsilon^{\prime}}
\def\vep{\varepsilon}
\def\ra{\rightarrow}
\def\ppg{\pi^+\pi^-\gamma}
\def\vp{{\bf p}}
\def\ko{K^0}
\def\kb{\bar{K^0}}
\def\al{\alpha}
\def\ab{\bar{\alpha}}
\def\be{\begin{equation}}
\def\ee{\end{equation}}
\def\beq{\begin{equation}}
\def\eeq{\end{equation}}
\def\bea{\begin{eqnarray}}
\def\eea{\end{eqnarray}}
\def\CPbar{\hbox{{\rm CP}\hskip-1.80em{/}}}
\newcommand{\lsim}{
\mathrel{\hbox{\rlap{\hbox{\lower4pt\hbox{$\sim$}}}\hbox{$<$}}}}

\thispagestyle{plain}

\def\bib{B\kern-.05em{I}\kern-.025em{B}\kern-.08em}
\def\btex{B\kern-.05em{I}\kern-.025em{B}\kern-.08em\TeX}

\title{Progress on semi-leptonic  $B_{(s)}$ decays}

\markboth{}{}

\author{\footnotesize GIULIA RICCIARDI}

\address{Dipartimento di Fisica, Universit\`a  di Napoli Federico II \\
Complesso Universitario di Monte Sant'Angelo, Via Cintia,
I-80126 Napoli, Italy\\
and \\
 INFN, Sezione di Napoli\\
Complesso Universitario di Monte Sant'Angelo, Via Cintia,
I-80126 Napoli, Italy\\
giulia.ricciardi@na.infn.it}

\maketitle

\begin{abstract}
We  report on the current status of   semi-leptonic $B_{(s)}$ decays, including rare decays,  and the extraction of the absolute values of the  CKM  matrix elements  $|V_{cb}|$ and $|V_{ub}|$.

\end{abstract}
\section{Introduction}

The  values of the CKM matrix   elements  are not
predictable within the Standard Model (SM) and  have to be inferred by experimental data.
At present, the most precise values of  $|V_{cb}|$ and $|V_{ub}|$ come from semi-leptonic decays,  driven, at parton level, by the tree level decays $ b \to c (u)  l \nu$.
%
%
 The  determinations of $|V_{cb}|$ and $|V_{ub}|$  from inclusive and exclusive decays rely on
different theoretical calculations, each with different (independent) uncertainties, and on
different experimental techniques which have, to a large extent, uncorrelated
statistical and systematic uncertainties. This independence makes
the comparison of $|V_{cb}|$ and $|V_{ub}|$ values from inclusive and exclusive decays an interesting test of our physical understanding.
An indirect determination of
 $|V_{ub}|$ is also given by
the measurement of the rate of the  leptonic decays $ B^+  \rightarrow l^+ \nu $,
provided that the $B$-decay constant is known from theory. This determination is disadvantaged  by the helicity suppression and the possibility of a more relevant role of new physics.
%
%
We summarize significant and recent results on heavy-to-heavy and heavy-to-light semi-leptonic decays, as well as the actual scenario of $|V_{cb}|$ and $|V_{ub}|$ extraction.
We also discuss $B$ meson semi-leptonic decays to excited states of the charm meson spectrum. Lastly,  we outline the status of  rare decays, where hints of new physics have been very recently reported.
Reviews on semi-leptonic $B$ decays and on CKM matrix elements extraction are  already available\footnote{see e.g.  \refcite{Ricciardi:2012pf}, \refcite{Ricciardi:2012dj}, \refcite{Ricciardi:2013jf} and references therein.}, but an update  seems timely, since  progress in  data and theory is quickly accumulating.


\subsection{Exclusive heavy-to-heavy  decays}
\label{subsectionExclusive decays}

In $B$ decays, approximations and techniques of the  heavy quark effective theory (HQET) are used, due to the large $b$-quark mass.
In  $B \to D^{(\ast)}$ semi-leptonic decays, also the
mass  of the $c$-quark can be considered  large compared to the QCD scale, allowing further approximations.

The form factors depend on  $ \omega= v_B \cdot v_{D^{(\ast)}}$, the only scalar formed from the $B$ and $D^{(\ast)}$ velocities ($v_B^2= v_{D^{(\ast)}}^2=1$ by definition). The scalar $\omega$
 is related to  $q^2$,  the momentum transferred
to the lepton pair, according to the relation
$ \omega= (m_B^2+m_{D^{(\ast)}}^2 -q^2)/(2 m_B m_{D^{(\ast)}})$.
For negligible lepton masses ($l=e, \mu)$,
the  differential ratios
can be written as
\bea
\frac{d\Gamma}{d \omega} (B \rightarrow D\,l {\nu})  &=&  \frac{G_F^2}{48 \pi^3}\,   (m_B+m_D)^2  m_D^3 \,
(\omega^2-1)^{\frac{3}{2}}\,  |V_{cb}|^2 {\cal G}^2(\omega)\nonumber \\
\qquad\frac{d\Gamma}{d \omega}(B \rightarrow D^\ast\,l {\nu})
&=&  \frac{G_F^2}{48 \pi^3}  (m_B-m_{D^\ast})^2 m_{D^\ast}^3 \chi (\omega)  (\omega^2-1)^{\frac{1}{2}} |V_{cb}|^2  {\cal F}^2(\omega)
 \label{diffrat}
\eea
in terms of a single form factor ${\cal G}(\omega)$ and ${\cal F}(\omega)$, for $B \to D l  \nu$ and $B \to D^{\ast} l  \nu$, respectively. The function  $\chi (\omega)$ is a phase space
factor
\beq
\chi (\omega) \equiv  (\omega+1)^2 \left[ 1+\frac{4 \omega \, (m_B^2 - 2 \omega\, m_B m_{D^\ast}+m_{D^\ast}^2)}{(\omega +1)\,    (m_B-m_{D^\ast})^2} \right]
\eeq
In the HQET limit,  both the form factors become \cite{Isgur:1989vq,Isgur:1989ed}
\beq
{\cal G}(1) = {\cal F}(1) =1
\eeq
at the zero recoil point $\omega=1$; when $D^{(\ast)}$ is at rest with respect to $B$, the light constituents of the initial and final hadrons are not affected
by the transition $b \to c$.
For finite values of $m_b$ and $m_c$, these unity values are altered by perturbative QCD and EW corrections, and by order $1/m_{c,b}^n$  non-perturbative corrections.
Schematically
\beq
{\cal F}(1) = \eta_{EW} \eta_A ( 1 + \delta_{1/m^2}+ \dots)
\label{schemff}
\eeq
where $ \delta_{1/m^2}$ are power corrections  which are suppressed by a factor of at least $\Lambda_{QCD}^2/m_c^2 \sim 3 \%$,
$\eta_{EW}$ is
the electroweak enhancement factor  \cite{Sirlin:1981ie} and $\eta_A(\alpha_s) $ is a short distance QCD coefficient known at order $\alpha_s^2$ \cite{Czarnecki:1996gu, Czarnecki:1997hc,Czarnecki:1998kt }.
For ${\cal F}$ non-perturbative linear corrections are absent at zero recoil \cite{Luke:1990eg}  and the leading terms are quadratic in $1/m_{b,c}$.
A  relation similar to Eq. \eqref{schemff} holds for ${\cal G} (1)$,  with the addition of  linear  corrections $ \delta_{1/m}$.

For the determination of  $|V_{cb}|$, the decay $ \bar B \to D^\ast l \bar \nu $ is generally preferred,  because of higher experimental  rate, and the absence of  nonperturbative  linear corrections to the form factor  ${\cal F}$ at zero recoil.
The starting point is the experimental fit of  the product  $|{\cal F}(\omega) V_{cb}|$, taken at $\omega
\neq 1$ to avoid the kinematic suppression
factors.
Instead, the theoretical evaluation of the form factor ${\cal F}(\omega)$
is, in most cases, performed  at  $\omega
= 1$,  where relation \eqref{schemff} holds. This mismatch introduces a dependence on the
extrapolation from $\omega \neq 1$ to $\omega =1$.

 The most recent  Heavy Flavor Averaging
Group  (HFAG) experimental  fit \cite{Amhis:2012bh} gives
\beq  |V_{cb}\, {\cal F}(1)| = (35.85 \pm 0.11 \pm 0.44 ) \times   10^{-3} \label{VcbexpF1} \eeq
%
%
%
%
assuming a  form factor parametrization devised in 1998 \cite{Caprini:1997mu}, but  rescaled to  more recent  parameter values.
The Belle  measurement, with 711 ${\mathrm{fb}}^{-1 }$ of data collected,  gives currently the most precise values  \cite{Dungel:2010uk}, followed by the BaBar global fits \cite{Aubert:2007rs}, with  results in agreement.

The FNAL/MILC  collaboration has  performed
the  non perturbative determination  of the form factor ${\cal F}(1)$
in the lattice unquenched approximation,  which includes loops of up, down and
strange sea quarks \cite{Bernard:2008dn, Bailey:2010gb}. The up and down quarks are usually
taken to be degenerate, so those simulations are referred to as $n_f= 2+1$.
 There is a very recent update that uses
the full suite of MILC (2+1)-flavor asqtad ensembles with lattice spacings as small as 0.045 fm and light-to-strange-quark mass ratios as low as 1/20 \cite{Bailey:2014tva}, giving
\beq  {\cal F}(1)
=0.906\pm 0.004 \pm  0.012  \label{VcbexpF2}  \eeq
without the EW enhancement factor $\eta_{EW}$.
The first error is statistical and the second one systematic.
The estimate for $|V_{cb}| $, using the latest HFAG average, is reported in Table \eqref{phidectab2}.
The QCD error is now commensurate with the experimental error. At the current level of precision, it would be important to extend
the calculation to nonzero recoil. Indeed, at finite momentum transfer,  only old  quenched lattice results are  available \cite{deDivitiis:2008df}; which,
 combined with 2008 BaBar data \cite{Aubert:2007rs},  give the $|V_{cb}| $ value also reported in table \eqref{phidectab2}.

The lattice calculations have to be compared with non-lattice ones. The more recent value obtained
by  using zero recoil sum rules reads
 \cite{Gambino:2010bp, Gambino:2012rd}
\beq {\cal F}(1) = 0.86 \pm 0.02 \label{gmu} \eeq
Full $\alpha_s$ and up to $1/m^3$ corrections have been included; the impact of $1/m^4$ and $1/m^5$ corrections has been estimated.
A recent parameter update related to power corrections does not affect significantly the results \cite{Gambino:2013rza}.
The corresponding estimate of $|V_{cb}| $, using the HFAG average in Eq.~(\ref{VcbexpF1}), is
reported in Table \eqref{phidectab2}.
The slightly smaller values for the form factors in sum rules determinations imply slightly higher values of $V_{cb}$.
The theoretical error  is more than twice the error in the lattice determinations.

Let us now consider $ B \rightarrow D \, l \, \nu$ decays.  The most recent  Heavy Flavor Averaging
Group  (HFAG) experimental  fit, adopting the same parametrization used for $ B \rightarrow D^\ast \, l \, \nu$ decays,  gives \cite{Amhis:2012bh}
\beq |V_{cb}\,  {\cal G}(1)| = (42.64 \pm 0.72 \pm 1.35)  \times
10^{-3} \label{BDfitdat} \eeq
Recent progress has been reported by the FNAL/MILC collaboration \cite{Qiu:2013ofa}, updating the values for the form factor at zero recoil in the unquenched form approximation, dating back to 2005-06 \cite{Okamoto:2004xg, Laiho:2005ue}.
The propagating heavy quarks on the lattice
have been interpreted by means of an effective theory approach.
The hadronic
form factors have been computed  in the unquenched approximation as well.
The related estimate of $|V_{cb}| $  produced at  nonzero recoil   in the joint fit with the
 2009 Babar data \cite{Aubert:2009ac}, is
reported in Table \eqref{phidectab2}.
To quantify the improvement due to working at nonzero recoil,
 $|V_{cb}| $  is extracted by extrapolating the experimental data to zero recoil and compared
with the theoretical form factor at that point.
The result is found consistent with the  nonzero recoil determination,  with an error  25\% larger \cite{Qiu:2013ofa}.

The previous
   lattice determination of $|V_{cb}| $ at non-zero recoil was given in the quenched approximation \cite{de Divitiis:2007ui, deDivitiis:2007uk}.  It was  based on the step scaling method, which  avoids the recourse to HQET.
The related estimate of $|V_{cb}| $ by the Babar Collaboration  \cite{Aubert:2009ac},  with the same
 2009  data, gives an  higher value, reported in Table \eqref{phidectab2}
(the errors are  statistical, systematic and due to the theoretical uncertainty in the form factor $ {\cal G}$, respectively).

 The latest results from non-lattice calculations are about 10 years old,  and use the expansion around the ''BPS" limit, that is
the limit
where the parameters related to kinetic energy
and
the chromomagnetic moment are equal in the heavy quark expansion\cite{Uraltsev:2003ye}.
Under these assumptions, the Particle Data Group finds the form factor  \cite{Beringer:1900zz}
 \beq {\cal G}(1) =1.04 \pm  0.02 \eeq
and
the related
$|V_{cb}|$ value reported in  Table \eqref{phidectab2}.

\begin{table}[t]
\centering
\vskip 0.1 in
\tbl{ Exclusive $|V_{cb}|$  determinations}
{\begin{tabular}{l c} \hline \hline
\color{red}{{\it \bf Exclusive decay} } &  \color{red}{   $ \mathbf{ |V_{cb}| \times 10^{3}}$} \\
\hline
\hline
\color{blue}{$\bar{B}\rightarrow D^\ast \, l \, \bar{\nu}$  }
 \\
\hline
FNAL/MILC (Lattice unquenched) \cite{Bailey:2014tva}   & $ 39.04 \pm 0.49_{\mathrm{exp}} \pm 0.53_{\mathrm{QCD}} \pm 0.19_{\mathrm{QED}} $
\vspace{1mm} \\
\hline
HFAG (Lattice unquenched) \cite{Amhis:2012bh, Bernard:2008dn, Bailey:2010gb}   & $ 39.54 \pm 0.50_{\mathrm{exp}} \pm 0.74_{\mathrm{th}} $ \vspace{1mm}\\
\hline
Rome (Lattice quenched $\omega \neq 1$)  \cite{deDivitiis:2008df, Aubert:2007rs}   & $ 37.4 \pm 0.5_{\mathrm{exp}} \pm 0.8_{\mathrm{th}} $ \vspace{1mm}\\
\hline
HFAG (Sum Rules) \cite{Amhis:2012bh, Gambino:2010bp, Gambino:2012rd} & $   41.6\pm 0.6_{\mathrm{exp}}\pm 1.9_{\mathrm{th}} $  \vspace{1mm}\\
\hline
{\color{blue}{$  \bar{B}\rightarrow D \, l \, \bar{\nu} $}}
&   \\
\hline
FNAL/MILC  (Lattice unquenched $\omega \neq 1$)   \cite{Qiu:2013ofa, Aubert:2009ac}  & $38.5 \pm 1.9_{\mathrm{exp+lat}} \pm 0.2_{\mathrm{QED}} $  \vspace{1mm}\\
\hline
PDG (HQE + BPS)   \cite{Beringer:1900zz, Uraltsev:2003ye} & $ 40.6 \pm 1.5_{\mathrm{exp}} \pm 0.8_{\mathrm{th}} $  \vspace{1mm}\\
\hline
Rome (Lattice quenched $\omega \neq 1$) \cite{Aubert:2009ac, deDivitiis:2007ui} & $  41.6 \pm 1.8_{\mathrm{stat}}\pm 1.4_{\mathrm{syst}}
\pm 0.7_{\mathrm{FF}}  \label{lattunq} $ \vspace{1mm} \\
\hline
\hline
\end{tabular}}
\label{phidectab2}
\end{table}

Data on $ B^0 \to D^{(\ast) +} \mu^- \, \nu$  decays have been  provided until now by electron-positron machines, especially  the dedicated $B$-Factories BaBar and Belle.
At LHCb, statistics is accumulating and about 5 million $ B^0 \to D^{(\ast) +} \mu^- \, \nu$ decays are available \cite{Tilburg}. At   hadron colliders semi-leptonic decays have an high branching ratio, and muons in the final state allow an easy triggering. However, the
measurements of $|V_{cb}|$ and $|V_{ub}|$ imply the reconstruction, in the $b$-hadron rest frame, of observables difficult to measure at hadron colliders,  such as the
squared invariant mass of the lepton pair $q^2$.
At LHCb,  it is possible to improve the $q^2$ resolution
 by exploiting the separation between primary and secondary
vertices, determining the $B$ flight direction vector and measuring
the neutrino momentum with a two-fold ambiguity \cite{Bozzi:2013aba}.

The   $ B \to D^{(\ast)} \tau  \nu_\tau$ decays are more difficult to measure,
since  decays into the heaviest $\tau$ lepton are suppressed and there are
 multiple neutrinos in the final state, following the $\tau$ decay.
Multiple neutrinos stand in the way of the reconstruction of the invariant mass of $B$ meson, and additional constraints related to the $B$ production are required. At the $B$ factories, a major constraint exploited is the fact that $B$ mesons are produced from the process $e^+e^- \to \Upsilon (4 S)\to B \bar B$.

The latest results belong to the BaBar Collaboration, that has measured
the $\xbar{B} \to D^{(\ast)} \tau^- \xbar{\nu}_\tau$  branching fractions normalized to the
corresponding $\xbar{B} \to D^{(\ast)} l^- \xbar{\nu}_l$ modes (with $l=e , \mu$) by using  the full BaBar data sample, and found \cite{Lees:2012xj, Lees:2013uzd}
\bea
{\cal{R}}^\ast_{\tau/l} &\equiv&  \frac{{\cal{B}}( \xbar{B} \to D^\ast \tau^- \xbar{\nu}_\tau)}{{\cal{B}}( \xbar{B} \to D^\ast l^- \xbar{\nu}_l)}= 0.332 \pm 0.024 \pm 0.018 \nonumber \\
{\cal{R}}_{\tau/l} &\equiv&  \frac{{\cal{B}}( \xbar{B} \to D \tau^- \xbar{\nu}_\tau)}{{\cal{B}}( \xbar{B} \to D l^- \xbar{\nu}_l)}= 0.440 \pm 0.058 \pm 0.042
\label{ratiotau}
\eea
where the first uncertainty is statistical and the second is
systematic.
The results  exceed the SM predictions ${\cal{R}}^\ast_{\tau/l} (SM)= 0.252\pm 0.003$ and
 ${\cal{R}}_{\tau/l} (SM)= 0.297\pm 0.017$ by $2.7 \sigma$ and $2.0 \sigma$, respectively.
The combined significance of this disagreement is $3.4\sigma$ \cite{Lees:2012xj, Lees:2013uzd}.
Other estimates give
${\cal{R}}_{\tau/l} (SM) = 0.31 \pm 0.02$, with a combined phenomenological and lattice analysis  \cite{Becirevic:2012jf}, and  a similar result,  ${\cal{R}}_{\tau/l} (SM) = 0.316 \pm 0.012 \pm 0.007$, where the errors are statistical and total systematic, respectively, is found
in  a  (2+1)-flavor lattice QCD calculation
\cite{Bailey:2012jg}.
Both analysis reduce the significance of the discrepancy for ${\cal{R}}_{\tau/l}$
below $2\sigma$.

The BaBar results \eqref{ratiotau} are in agreement (with smaller uncertainties)   with measurements by
 Belle using the $\Upsilon(4S)$ data set that
corresponds to an integrated luminosity of 605 $\mathrm{fb}^{-1}$ and  contains
 $657  \times 10^6$  $B \xbar{B}$ events \cite{Adachi:2009qg}.
The branching ratio measured values have
consistently exceeded the SM expectations with increased precision, that starts to be enough to constrain NP.
Latest data from BaBar are not compatible with a
charged Higgs boson in the type II two-Higgs-doublet model
and with large portions of the more general type III two-Higgs-doublet model \cite{Lees:2013uzd}.
%
%
Several  NP frameworks have been studied  that  explain (or fail to) this alleged  hint of breaking of lepton-flavour universality.
Minimal flavor violating models,
 right-right vector and right-left scalar quark currents,  leptoquarks, quark and lepton
compositeness models have been investigated \cite{Fajfer:2012jt, Sakaki:2013bfa}, modified couplings  \cite{Abada:2013aba, Datta:2012qk},  additional tensor operators  \cite{Biancofiore:2013ki}, charged scalar contributions \cite{Dorsner:2013tla},  aligned  two Higgs doublet models \cite{Celis:2012dk},
  effective Lagrangians
\cite{Fajfer:2012vx, Datta:2012qk}, new sources of CP violation\cite{Hagiwara:2014tsa}, and so on.

There is room for improvement in current statistic limits for measurements of
 ${\cal{R}}_{\tau/l}$. It would be interesting to investigate if  the results of the Belle analysis shift  towards the SM results, obtained by  global unitarity-triangle fits,  by
using the full $\Upsilon(4S)$ data
sample containing $772 \times 10^6$  $B \xbar{B}$ pairs and the  improved hadronic tagging, as happened in  the case of  purely
 leptonic decays $B^- \to \tau^- \xbar{\nu}_\tau$  \cite{Adachi:2012mm}.

The Belle-II
experiment
will have an integrated luminosity over 30 times greater than that of the combined BaBar
and Belle datasets. It should start in 2016, and in the next decade provide accurate measurements to investigate possible NP contributions.
The estimate error with 75 $\mathrm{ab}^{-1}$ at the $\Upsilon(4S)$ is around 1\%.

In this section, we have always implicitly alluded to $B$ decays, but  semi-leptonic $B_s$ decays can also  probe CKM matrix elements.
Moreover, semi-leptonic $B^0_s$
decays are used as a normalization mode for various
searches for new physics at hadron colliders and at Belle-II.
The  presence of the heavier spectator strange quark is bound to introduce  some amount of SU(3) symmetry
breaking, which may  affect width ratios \cite{Gronau:2010if, Bigi:2011gf, Fan:2013kqa} of the kind of
\beq \frac{\Gamma(B_s \to X l \nu )}{\Gamma(B_d \to X l \nu )} \eeq
The branching fractions of  $B_s \to X l \nu$ decays, where $X$ is an arbitrary final state, have been measured  at BaBar \cite{Lees:2011ji} and Belle \cite{Oswald:2012yx},
in  data-sets obtained from  energy scans above the $\Upsilon(4S)$.
Belle has profited of the large available
 data sample of 121 $\mathrm{fb}^{-1}$ near the $\Upsilon(5S)$ resonance to perform the
most precise measurement of the
branching fraction \cite{Oswald:2012yx}
\beq
{\cal B}(B_s \to X l \nu) = [10.6 \pm 0.5_{\mathrm{stat}}   \pm 0.7_{\mathrm{syst}}  ]\%
\eeq
Other measurements of $B_s$  branching ratios have also been reported, precisely
of $B_s^0 \to D_{s1}^- (2536) \mu^+ \nu X$ decays  by \D0 \cite{Abazov:2007wg} and of
$\xbar{B}_s^0 \to D^{\ast +}_{s2} (2573) \mu^- \xbar{\nu} X$ decays  by LHCb \cite{Aaij:2011ju}.

Outside the $\Upsilon(4S)$ region,
the determination of $|V_{cb}|$ is also possible   by studying the baryonic  $\Lambda_b^0 \to \Lambda_c^+ l^- \xbar{\nu}$ decays \cite{Stone:2014mza}.

\section{$B$-Mesons Decays to Excited $D$-Meson States}


The increased interest in
semi-leptonic $B$ decays to excited states of the
charm meson spectrum  derives  by the fact that they
contribute
as a background to the direct decay $ B^0 \to D^{\ast } l  \nu$ at the B factories, and, as a consequence, as
 a source of systematic error in the $|V_{cb}|$ measurements.

The spectrum of mesons consisting of a charm and an
up or a down anti-quark is poorly known.  In the non-relativistic constituent quark model,
the open charm system  can be classified according to  the radial quantum
number and to the eigenvalue $L$ of the relative angular momentum
between  the c-quark and the light degrees of freedom,
The low-mass spectrum is comprised of the
ground states ($1S$, $L=0$), that is $D$ and $D^\star$ mesons,  the orbital excitations with angular
momentum $L =1,2$ ($1P$, $1D$), and the first radial excitations ($2S$).

The four states with $L=1$ are generically
denoted as $D^{\ast \ast}$\footnote{Sometimes in literature this term is extended to include all particles in the low-mass spectrum except the ground states.}.
and have been identified as
  $D^\ast_0(2400)$,  $D_1(2420)$, $D^\prime_1(2430)$ (or $ D_1(2430)$), and $D^\ast_2(2460)$\cite{Beringer:1900zz}.
Two of them,  $D_1(2420)$ and $D^\ast_2(2460)$,
 have relatively narrow widths, about 20-30 MeV, and have been observed and studied  by a number of experiments
since the nineties (see Ref. \refcite{Aubert:2009wg} and Refs. therein).
The  other
two  states,  $D^\ast_0(2400)$, $D_1^\prime(2430)$,  are more difficult to detect due to the large width, about 200-400 MeV, and their observation has started more recently
\cite{Abe:2003zm, Abazov:2005ga, Abdallah:2005cx,Aubert:2008ea,Liventsev:2007rb}
%

%

In 2010 BaBar has observed,  for the first time, candidates for the radial excitations of the $D^0$, $D^{\ast 0}$ and $D^{\ast +}$, as well as the $L=2$ excited states of the $D^0$ and $D^+$ \cite{delAmoSanchez:2010vq}.
Resonances in the $2.4$-$2.8$  ${\mathrm{GeV/c}}^2$ region  of hadronic masses have also  been identified at LHCb
\cite{Aaij:2013sza}.
In the same region,  data
are available on
 semi-leptonic $B$ decays to final states containing a $D_s^{(\ast)+} K$ system
\cite{delAmoSanchez:2010pa, Stypula:2012mf}.


%
%
The not completely clear experimental situation is mirrored by two theoretical puzzles.
Most   calculations, using sum rules \cite{LeYaouanc:1996bd,Uraltsev:2000ce}, quark models \cite{Morenas:1997nk,  Ebert:1998km, Ebert:1999ga},  OPE \cite{Leibovich:1997em, Bigi:2007qp} (but not   constituent quark models \cite{Segovia:2011dg}),
indicate that  the narrow width states dominate over
the broad  $D^{\ast\ast}$ states, in contrast to experiments (the ``1/2 vs 3/2 puzzle").
One possible  weakness common to these theoretical approaches is that they are derived in the heavy quark limit and
corrections might
be  large.
The other puzzle   is that
 the sum of the measured semi-leptonic exclusive rates having $D^{(\ast)}$ in the final state is less than
the inclusive one (``gap problem") \cite{Liventsev:2007rb, Aubert:2007qw}.
To overcome the difficulties to disentangle very broad resonances from continuum, both on theoretical and experimental sides,  it has been  suggested  to  clarify the comparison between theory and experiment analyzing   states analogous to $D_0^\ast$ and $D^\prime_1$, but narrow,
in particular  studying the decay  $B^0_s \rightarrow \bar D_{s J} \pi$
\cite{Becirevic:2012te}.
Other  theoretical suggestions to ease or solve the previous problems include  taking into account  an unexpectedly large $B$ decay
rate to the first radially excited $D^{\prime(\ast)} $
\cite{Bernlochner:2012bc, Gambino:2012rd}.
A recent proposal is to extract exclusive branching fractions
of semi-leptonic B-meson decays to charmed mesons from a fit to electron energy, hadronic mass and combined hadronic mass-energy momenta measured in inclusive $B \to X_c l \nu$ decays, as an alternative to direct measurements \cite{Bernlochner:2014dca}.

Recently, first dynamical lattice computations of the $\bar B \to D^{\ast \ast } l \nu$ form factors have been attempted, although  still preliminary \cite{Atoui:2013sca, Atoui:2013ksa}.

\subsection{Inclusive  $ B \rightarrow X_c \, l \, \nu_l$ decays}
\label{subsectionInclusive decays}

In  inclusive $ B \rightarrow X_c \, l \, \nu_l$ decays,  the final state
$X_c$ is an hadronic state originated by the charm  quark. There is no dependence on the details of the final state, and quark-hadron duality is generally assumed.
Long distance dynamics of the meson can be factorized by using an  OPE approach, which, combined with HQET,
gives to inclusive transition rates  the form  of an (Heavy Quark)  Expansion (HQE) in $1/m_b$.
The coefficients of the expansion are calculable  in perturbation theory, while
the hadronic
expectation values of the operators encode the
nonperturbative corrections and depend  on a   number of  HQE  parameters,
which  increase at increasing powers of $1/m_b$.
These parameters are  affected by the
 particular theoretical framework (scheme) that is
used to define the quark masses.
The HQE is  valid only for sufficiently inclusive measurements, therefore the relevant quantities to be measured are
global shape parameters (the first few moments of various kinematic distributions)
and the total rate.

At order $1/m_b^0$, the parton level, we have the usual $\alpha_s$ expansion,
which is known completely to order $\alpha_s$ and $\alpha_s^2$,
for the width and moments of the
lepton energy and hadronic mass distributions
 (see Refs. \refcite{Trott:2004xc}, \refcite{Aquila:2005hq}, \refcite{Pak:2008qt}, \refcite{Pak:2008cp}, \refcite{Biswas:2009rb}
and references therein). The terms of order $\alpha_s^{n+1} \beta_0^n$, where $\beta_0$ is the first coefficient of the QCD $\beta$ function, have also been computed following  the
 BLM procedure \cite{Benson:2003kp,Aquila:2005hq}.

At the next order in the  HQE, that is $ \Lambda_{QCD}^2/m_b^2$,
there are
two  operators, called the kinetic energy  and the chromomagnetic operator.
%
The perturbative corrections to the coefficient of the  kinetic matrix element  have been
 evaluated   at order $\alpha_s$  for generic observables, such as partial rates and moments \cite{Becher:2007tk, Alberti:2012dn}.
They lead to
numerically modest modifications of the width and moments.
Corrections at order $\alpha_s$  to the
coefficient of  the matrix element of the chromomagnetic operator  have also been   computed   \cite{ Ewerth:2009yr, Alberti:2013kxa}. The results have been employed to evaluate the
correction to the semi-leptonic decay width, the mean lepton energy, and the variance (second central moment)
of the lepton energy distribution.
The complete corrections of the $\alpha_s \Lambda_{QCD}^2/m_b^2$ to the width  is a few per mill on the width, but the corrections to the first two leptonic moments  are of the same
order of the experimental errors \cite{Alberti:2013kxa}. The  estimate of these effects on  the determination of $|V_{cb}|$ is under the way.

%
%
Neglecting  perturbative corrections, i.e. working at tree level,  contributions to various observables   have been
computed at order $1/m_b^3$ and estimated at order $1/m_b^{4,5}$,
 $1/m^3_b \, m^2_c$; contributions at order
$\alpha_s(m_c)1/m^3_b\, m_c$ and the so-called intrinsic charm term
 have been estimated as well\cite{Gremm:1996df, Dassinger:2006md, Bigi:2005bh, Breidenbach:2008ua,  Bigi:2009ym, Mannel:2010wj}.

\begin{table}[t]
\centering
\vskip 0.1 in
\tbl{ $|V_{cb}|$ inclusive determinations}
{\begin{tabular}{l c} \hline \hline
\color{red}{{\it \bf Inclusive decays} } &  \color{red}{   $ \mathbf{ |V_{cb}| \times 10^{3}}$} \vspace{1mm} \\
\hline
\hline
global fit, kin scheme, $m_c$ constraint (HFAG) \cite{Amhis:2012bh}  & $41.88 \pm 0.44_{\mathrm{fit}} \pm 0.59_{\mathrm{th}}  $
\vspace{1mm} \\
\hline
global fit, kin scheme,  $ B \to
X_s \gamma$  constraint (HFAG)  \cite{Amhis:2012bh} & $41.94 \pm 0.43_{\mathrm{fit}} \pm 0.59_{\mathrm{th}} $
\vspace{1mm} \\
\hline
global fit, kin scheme,   $m_c$ constraint  \cite{Gambino:2013rza}   & $42.42 \pm 0.86$
\vspace{1mm} \\
\hline
\hline
\end{tabular}}
\label{phidectab02}
\end{table}

A global fit   is a simultaneous fit to
 HQE  parameters, quark masses and absolute values of  CKM matrix elements obtained by  measuring
 spectra plus all
available moments.
Only the HQE parameters associated with  $O(1/m^{2,3}_b)$  corrections are
routinely fitted from experiments, one reason being the growth in their number at higher orders.
The
HFAG global fit  employs as  experimental inputs  the (truncated) moments of the
lepton energy $E_l$  (in the $B$ rest frame) and the $m_X^2$  spectra in $B \to X_c l \nu$ \cite{Amhis:2012bh}.
The results, although sufficient
for determining  $|V_{cb}|$,  measure the $b$-quark mass only to about 50 MeV precision. To get higher precision,
additional constraints are introduced:  the photon energy moments in
 $ B \to
X_s \gamma$,  or a precise  constraint  on the $c$-quark  mass.
The actual HFAG  global fit is performed in the kinetic scheme and  yields
  \beq |V_{cb}| = (41.88 \pm 0.44_{\mathrm{fit}} \pm 0.59_{\mathrm{th}}) \times 10^{-3}\eeq
with the constrained  value
$m_c^{\overline{\mathrm{MS}}} \mathrm{(3 GeV)} = (0.998 \pm 0.029)$ GeV, obtained using low-energy sum rules \cite{Dehnadi:2011gc}.
By using the
 $ B \to
X_s \gamma$ constraints, it gives
\beq |V_{cb}| = (41.94 \pm 0.43_{\mathrm{fit}} \pm 0.59_{\mathrm{th}}  ) \times 10^{-3}\eeq
The precision is
higher than in the exclusive determinations, being  about 1.7\%.
Another recent determination in the kinetic scheme  of $|V_{cb}|$ gives \cite{Gambino:2013rza}
\beq |V_{cb}| = (42.42 \pm 0.86 ) \times 10^{-3}\eeq
In analogy to the HFAG determination, it uses the full $O(\alpha_s^2/m_b^0)$ calculation and  no $O(\alpha_s/m_b^2)$ calculations, but it employs  a slightly different constraint on $m_c$. The error is also slight larger, about 2\%, comparable with the best errors of the exclusive determination.
The results have been collected in Table \ref{phidectab02}, that, compared with the results in Table \ref{phidectab2} for the exclusive determination, shows a tension in most cases around 2-3$\sigma$, the precise number depending on the values used for the comparison.
One could also compare with indirect fits, provided  by the UTfit collaboration \cite{Utfit}
\beq  |V_{cb}|  = (42.12 \pm  0.07) \times 10^{-3} \eeq
   and  by the CKMfitter collaboration (at $1 \sigma$) \cite{CKMfitter}.
\beq |V_{cb}|  = (41.51^{+0.56}_{-1.15}) \times 10^{-3} \eeq
Indirect fits  prefer a value for $|V_{cb}|$ that is closer to the (higher)
inclusive determination.

High statistic $B$-factories have greatly contributed to the increase in measurement precision with respect to previous experiments, and the high statistics at  Belle II at SuperKEKB is expected, within  the next decade,  to push errors on $|V_{cb}|$ down to 1\% \cite{Yashchenko}.

\subsection{Exclusive  heavy-to-light decays}
\label{Exclusivesemi-leptonicdecays33}

The analysis of
exclusive charmless semi-leptonic decays, in particular the  $\bar B \rightarrow \pi l \bar \nu_l$ decay,   is currently employed to determine the CKM parameter $|V_{ub}|$.
The  $ B \rightarrow \pi l  \nu$  decays depend on a single form factor $f_+(q^2)$,
 in the limit of zero leptonic masses.
%
%
The  first lattice determinations of  $f_+(q^2)$  based on unquenched  simulations have been obtained by  the HPQCD collaboration
\cite{Dalgic:2006dt} and  the Fermilab/MILC collaboration\cite{Bailey:2008wp}; they are  in substantial agreement.
These analyses, at $q^2 > 16$ $\mathrm{GeV}^2$, together with
latest data  on   $ B \rightarrow \pi l  \nu$ decays coming from Belle and BaBar, and 2007 data from CLEO,   have been employed  in the actual  HFAG averages \cite{Amhis:2012bh}.
The results have been reported in Table \ref{phidectab03}. Also,  HFAG has performed a simultaneous
fit of the BCL parametrization \cite{Bourrely:2008za} to  lattice results  and
experimental data, to exploit all the available
information  in the full $q^2$ range,
which has given  the following average value
\beq |V_{ub}|  = (3.28 \pm  0.29) \times 10^{-3}\eeq
%

\begin{table}[t]
\centering
\vskip 0.1 in
\tbl{$|V_{ub}|$  exclusive determinations}
{\begin{tabular}{l c} \hline \hline
\color{red}{{\it \bf Exclusive decays} } &  \color{red}{   $ \mathbf{ |V_{ub}| \times 10^{3}}$} \vspace{1mm} \\
\hline
\hline
\hline
\color{blue}{$\bar B \rightarrow \pi l \bar \nu_l$    }
 \\
\hline
HPQCD ($q^2 > 16 $) (HFAG)  \cite{Dalgic:2006dt, Amhis:2012bh}   & $ 3.52 \pm 0.08^{0.61}_{0.40}  $
\vspace{1mm} \\
\hline
 Fermilab/MILC ($q^2 > 16 $)  (HFAG)  \cite{Bailey:2008wp, Amhis:2012bh} & $ 3.36 \pm 0.08^{0.37}_{0.31}  $
\vspace{1mm} \\
\hline
lattice, full $q^2$ range   (HFAG)   \cite{Amhis:2012bh}   & $3.28 \pm 0.29$
\vspace{1mm} \\
\hline
LCSR  ($q^2 < 12 $)  (HFAG)   \cite{Khodjamirian:2011ub, Amhis:2012bh}   & $3.41 \pm 0.06^{+0.37}_{-0.32}$
\vspace{1mm} \\
\hline
LCSR  ($q^2 < 16 $)  (HFAG)   \cite{Ball:2004ye, Amhis:2012bh}   & $3.58 \pm 0.06^{+0.59}_{-0.40}$
\vspace{1mm} \\
\hline
lattice+ LCSR  (Belle)   \cite{Sibidanov:2013rkk}   & $3.52 \pm 0.29$
\vspace{1mm} \\
\hline
\hline
\end{tabular}}
\label{phidectab03}
\end{table}

On the lattice front,  the  Fermilab/MILC collaboration has recently presented an update, a blinded analysis
over a range of lattice spacings $a \sim 0.045$-$0.12$ fm \cite{Du:2013kea}.
Preliminary results have been presented by the  ALPHA \cite{Bahr:2012vt, Bahr:2012qs} ($n_f=2$) HPQCD \cite{Bouchard:2012tb}
($nf=  2 + 1$), and the RBC/UKQCD \cite{Kawanai:2012id} ($nf=  2 + 1$)  Collaborations.
Recent results are also available on a fine lattice (lattice spacing $a \sim 0.04$ fm) in the quenched approximations by the QCDSF collaboration \cite{AlHaydari:2009zr}.

In the complementary   kinematic region, at
large recoil,
 direct LCSR calculations of the semi-leptonic  form factors  are available, which have benefited by   progress in pion distribution amplitudes, next-to-leading (NLO) and leading (LO) higher order twists (see e.g.~\cite{Khodjamirian:2011ub, Bharucha:2012wy,Li:2012gr} and Refs. within). The $|V_{ub}|$ estimate are generally higher than the corresponding lattice ones, but still in agreement, within the relatively larger theoretical errors.
The estimated values for $|V_{ub}|$
according to  LCSR \cite{Ball:2004ye, Khodjamirian:2011ub}
provided by HFAG have been reported  in Table \ref{phidectab03}.
Higher values for $|V_{ub}|$ have been computed  in the relativistic quark model \cite{Faustov:2014zva}.

A recent analysis using hadronic reconstruction by Belle  \cite{Sibidanov:2013rkk} leads to a branching ratio of
$ {\cal{B}} ( B^0 \to \pi^- l^+ \nu) = (1.49 \pm 0.09_{\mathrm{stat}} \pm 0.07_{syst}) \times 10^{-4}$,
which is competitive with the more precise
results from untagged measurements.
From the $ \bar B \to \pi  l^- \bar \nu_l$
decay, Belle extracts the value
\beq |V_{ub}|  = (3.52 \pm  0.29) \times 10^{-3}\eeq
using
their measured partial branching fractions, and combining    LCSR, lattice points and the BCL \cite{Bourrely:2008za},
 description of the
$f_+(q^2)$ hadronic form factor \cite{Sibidanov:2013rkk}. This value
 is also reported in Table  \ref{phidectab03}.

Recently, significantly improved branching ratios of other
 heavy-to-light semi-leptonic decays have been reported, that reflect on
increased precision
for $|V_{ub}|$ values inferred by these decays.
$|V_{ub}|$ has been extracted from  $ B^+ \rightarrow \omega l^+ \nu$  \cite{Lees:2013gja},  yielding.  with the LCSR form factor determination \cite{Ball:2004rg}
\beq
|V_{ub}|  =
(3.41\pm 0.31 ) \times 10^{-3}  \eeq
and, with the ISGW2 quark model\cite{ Scora:1995ty}
\beq
|V_{ub}|  =
(3.43\pm 0.31 ) \times 10^{-3}  \eeq
A major problem is that the quoted uncertainty does not include any uncertainty from theory, since uncertainty estimates of the
form-factor integrals are not available.

Other channels to have been investigated are $ B \to \rho l \nu$ decays.
By comparing the
measured distribution in $q^2$, with an upper limit at $q^2 = 16$ GeV, for  $ B \to \rho l \nu$ decays,
with LCSR  predictions for the form factors\cite{Ball:2004rg},
 the $|V_{ub}|$ value reads  \cite{delAmoSanchez:2010af}
\beq
|V_{ub}|  =
(2.75 \pm 0.24 ) \times 10^{-3}  \eeq
and with the ISGW2 quark model\cite{ Scora:1995ty}.
\beq
|V_{ub}|  =
(2.83 \pm 0.24 ) \times 10^{-3}  \eeq
More recent results have been presented by a Belle tagged analysis  \cite{Sibidanov:2013rkk},
resulting in a  branching fraction
$ {\cal{B}} ( \bar B^0 \to \rho^+ l^- \bar \nu_l) = (3.34 \pm0.23) \times 10^{-4}$,
which is 43\% (2.7$\sigma$) higher
than the current Particle Data Group  value $ {\cal{B}}^{PDG} ( \bar B \to \rho l^- \bar \nu_l) = (2.34 \pm 0.15 \pm 0.24) \times 10^{-4}$
\cite{Beringer:1900zz}  and has a better precision (almost a factor  two).
In the same analysis \cite{Sibidanov:2013rkk},   an evidence of a broad resonance around 1.3
GeV dominated by the
$B^+\to f_2 l \nu$
decay has also been reported, for the first time.

Other interesting  channels  are
 $ B \rightarrow \eta^{(\prime)} l \nu $ decays, not yet sufficiently
precise  to be  used for
$|V_{ub}|$ extraction.
The
value of the ratio
\beq \frac{{\cal{B} }(B^+ \to \eta^\prime l^+ \nu_l)}{{\cal{B} }(B^+ \to \eta l^+ \nu_l)}=0.67 \pm 0.24_{\mathrm{stat}} \pm 0.11_{\mathrm{syst}} \eeq
seems to
allow an important gluonic singlet contribution to the
$\eta^\prime$
form factor \cite{delAmoSanchez:2010zd, DiDonato:2011kr}.

In future prospects, other channels that can be valuable to extract $|V_{ub}|$ are $B_s \to K^{(\ast)} l \xbar{\nu}$ decays \cite{Meissner:2013pba}. Other semileptonic decays, as  $B^- \to \pi^+\pi^- l^- \xbar{\nu}_l$, through the analysis of their angular variables,  can be used to measure dipion form factors \cite{Faller:2013dwa}.

Baryonic  semi-leptonic decays are the subject of a growing interest  \cite{Stone:2014mza}, in particular the  $|V_{ub}|$ sensitive $\Lambda^0_b \to p l^- \bar \nu$ decays, whose form factors have been evaluated  in the LCSR
\cite{Khodjamirian:2011jp} and in the lattice with static $b$-quarks frameworks \cite{Detmold:2013nia}.
Help in constraining the baryonic transition form factor in $B$ decays may instead come from the recent evidence for the semi-leptonic  decay $B^- \to p \bar p l^- \xbar{\nu}_l$ ($l=e,\mu$) \cite{Tien:2013nga}.

The  purely leptonic decay $B^- \to \tau^- \xbar{\nu}_\tau$, first  observed  by Belle in 2006 \cite{Ikado:2006un},
has the SM branching ratio
\beq {\cal{B}}(B^- \to \tau^- \xbar{\nu}_\tau) = \frac{G_F^2 m_B m_\tau^2}{8 \pi} \left(1- \frac{m_\tau^2}{m_B^2} \right)^2 f_B^2 |V_{ub}|^2 \tau_B
\eeq
 and its  measurement  provides a direct experimental determination of the
product  $f_B  |V_{ub}|$.
%
The new Belle result \cite{Adachi:2012mm}
\beq {\cal{B}}( B^- \to \tau^- \xbar{\nu}_\tau) =( 0.72^{+0.27}_{-0.25}\pm 0.11) \times 10^{-4}
\eeq
where the first errors are statistical and the second ones systematical, differs from previous BaBar and Belle results, which, averaged by HFAG  \cite{Amhis:2012bh}, give the branching fraction
\beq {\cal{B}}( B^- \to \tau^- \xbar{\nu}_\tau) =( 1.67 \pm 0.30) \times 10^{-4}
\eeq
By using the recent Belle value together with the previous Belle measurement based on
a semi-leptonic $B$
tagging method and taking into account
all the correlated systematic errors, the branching fraction is found to be \cite{Adachi:2012mm}
\beq {\cal{B}}( B^- \to \tau^- \xbar{\nu}_\tau) =( 0.96\pm 0.26) \times 10^{-4}
\eeq
Combining this value with the
 mean $B^+$-meson lifetime $\tau_B= 1.641 \pm 0.008$  \cite{Beringer:1900zz}
and their
average for the $B$ meson decay constant, $f_B=190.5 \pm 4.2$ MeV ($n_f=2+1$), the FLAG (Flavor Lattice Averaging Group) collaboration obtains \cite{Aoki:2013ldr}
\beq |V_{ub}|  = (3.87 \pm  0.52 \pm 0.09) \times 10^{-3}\eeq
where the first error comes from experiment and the second comes from the uncertainty
in $f_B$.
The accuracy is not yet enough to make this channel competitive for $
|V_{ub}|$ extraction. In contrast with previous experimental analyses, the new Belle data seem to indicate agreement with previous results from the SM.
Search of possible lepton flavour violations can also be made independent of $|V_{ub}|$
by building ratios of branching fractions, such as
\beq
R^\prime = \frac{\tau_{B^0}}{\tau_{B^+}} \frac{{\cal{B}}( B^+\to \tau^+ \nu_\tau) }
{{\cal{B}}( B^0 \to \pi^- l^+ \nu_l) }
\eeq

\subsection{Inclusive  $ B \rightarrow X_u \, l \, \nu_l$  decays}
\label{vuninclusivo}

The extraction of $|V_{ub}|$ from inclusive decays requires to address theoretical issues absent in the inclusive $|V_{cb}|$ determination.
OPE techniques  are not applicable in the   so-called  endpoint or singularity or  threshold phase space region,
 corresponding to the kinematic region near the limits of
both the lepton energy  $E_l$ and $q^2$ phase space, where the rate is dominated by
the production of low mass final hadronic states.
This region is
plagued by the presence
 of large double (Sudakov-like)  perturbative  logarithms at all orders in the strong coupling.
Corrections  can be large  and need to be resummed at all orders \footnote{See e.g. ~ \refcite{DiGiustino:2011jn,Aglietti:2007bp,Aglietti:2005eq,Aglietti:2005bm}, \refcite{Aglietti:2005mb}, \refcite{Aglietti:2002ew},\refcite{Aglietti:2000te}, \refcite{Aglietti:1999ur}, and references therein.}. The kinematics cuts due to the large $B \to X_c l \nu$ background enhance the weight of the threshold region with respect to the case of
  $b \rightarrow c$ semi-leptonic decays; moreover, in the latter, corrections are not expected  as singular as in the $ b \rightarrow u$ case, being  cutoff by the charm mass.

On the experimental side, efforts have been made
 to control the background and access to a large part of the phase space, so as to reduce,
on the whole,  the weight of the endpoint region.
 Latest results by Belle \cite{Urquijo:2009tp} and BaBar \cite{Lees:2011fv}
use their complete data sample, $ 657$ x $ 10^{6}$  $B$-$\xbar{B}$ pairs for Belle   and 467 x $ 10^{6}$ $B$-$\xbar{B}$ pairs for BaBar. Although the two analyses differ
in the treatment of the background, both collaborations claim to access $\sim  90$\% of the phase space.

On the theoretical side, several schemes are available. All of them are  tailored
to analyze data in the threshold region,  but  differ significantly
in their treatment of perturbative corrections and the
parametrization of non-perturbative effects.

The
average values for $|V_{ub}|$ have  been extracted  by HFAG   from the partial branching fractions, adopting
a specific theoretical framework and taking into account
 correlations among the various measurements
and theoretical uncertainties \cite{Amhis:2012bh}.
BaBar and Belle analysis, Refs. \refcite{Urquijo:2009tp} and  \refcite{Lees:2011fv}, as well as  the HFAG averages in Ref. \refcite{Amhis:2012bh}
rely on at least four different QCD calculations of the partial
decay rate: BLNP
by Bosch, Lange, Neubert, and Paz \cite{Lange:2005yw, Bosch:2004th, Bosch:2004cb}; DGE, the
dressed gluon exponentiation, by Andersen and Gardi \cite{Andersen:2005mj}; ADFR by Aglietti, Di Lodovico, Ferrara, and Ricciardi
\cite{Aglietti:2004fz, Aglietti:2006yb,  Aglietti:2007ik}; and GGOU by Gambino, Giordano, Ossola
and Uraltsev \cite{Gambino:2007rp}.
 These  QCD theoretical calculations are the ones taking into account  the whole set of experimental results, or most of it, starting from 2002 CLEO data \cite{Bornheim:2002du}.
They can be roughly  divided into approaches based on the estimation of the shape function (BLN, GGOUP) and on resummed perturbative QCD (DGE, ADFR).
Other theoretical schemes have been described  in Refs.  \refcite{Bauer:2001rc},\refcite{Leibovich:1999xf}, \refcite{Lange:2005qn}.
Although conceptually quite different, all the above approaches generally
lead to roughly consistent results when the same inputs are used and the
theoretical errors are taken into account.
The HFAG estimates \cite{Amhis:2012bh}, together with the latest estimates by BaBar and Belle
\cite{Urquijo:2009tp, Lees:2011fv}, are reported in Table \ref{phidectab3},

We can also compare with indirect fits
\beq
 |V_{ub}|  = (3.61 \pm  0.12) \times 10^{-3}
\eeq
by  UTfit \cite{Utfit}  and
\beq
 |V_{ub}|  = (3.49^{+0.21}_{-0.10}) \times 10^{-3}
\eeq
at $1 \sigma$ by CKMfitter \cite{CKMfitter}.
At variance
with the $|V_{cb}|$ case, the results of the global fit prefer a value for $|V_{ub}|$ that is closer to the (lower)
exclusive  determination.

\begin{table}[t]
\tbl{ $|V_{ub}|$ inclusive determinations}
{\begin{tabular}{l c c c c } \hline \hline
& \color{red}{{\it \bf Inclusive decays} }& \color{red}{  ( $ \mathbf{ |V_{ub}| \times 10^{3}}$ )}\vspace{1mm} \\
\hline
\hline
\hline
 & \color{blue}{BNLP   } \cite{Lange:2005yw, Bosch:2004th, Bosch:2004cb} &   \color{blue}{GGOU  }   \cite{Gambino:2007rp} &   \color{blue}{ADFR   } \cite{Aglietti:2004fz, Aglietti:2006yb,  Aglietti:2007ik} &   \color{blue}{DGE }   \cite{Andersen:2005mj}
 \\
\hline
BaBar   \cite{Lees:2011fv}&  $4.28 \pm 0.24^{+0.18}_{-0.20}  $  & $4.35 \pm 0.24^{+0.09}_{-0.10}  $  & $4.29 \pm 0.24^{+0.18}_{-0.19}  $  & $4.40 \pm 0.24^{+0.12}_{-0.13}  $
\vspace{1mm}  \\
\hline
 Belle  \cite{Urquijo:2009tp} &  $ 4.47 \pm 0.27^{+0.19}_{-0.21}  $  &  $4.54 \pm 0.27^{+0.10}_{-0.11}  $      & $4.48 \pm 0.30^{+0.19}_{-0.19}  $    & $4.60 \pm 0.27^{+0.11}_{-0.13}  $
\vspace{1mm} \\
\hline
HFAG  \refcite{Amhis:2012bh} &  $ 4.40 \pm 0.15^{+0.19}_{-0.21}  $ & $4.39 \pm  0.15^{ + 0.12}_ { -0.20} $ &  $4.03 \pm 0.13^{+ 0.18}_{- 0.12}$ &
$4.45 \pm 0.15^{+ 0.15}_{- 0.16}$
\vspace{1mm} \\
\hline
\hline
\end{tabular}}
\label{phidectab3}
\end{table}

\section{Rare decays}

The  increased luminosity of the actual experimental facilities and the possibility
 to explore rare decays in quantitative detail  have prompted  a lot of recent theoretical activity.

In the inclusive $B \rightarrow X_s l^+ l^-$ decays
the major theoretical uncertainties come from
the non-perturbative nature of the intermediate  $\bar c c $ states.
By cutting on the invariant di-lepton mass around the masses of the $J/\psi$  and  $\psi^\prime$  resonances,
rather precise determinations seem to be possible, since
below or above the   $\bar c c $  resonances,
the inclusive decay is dominated by perturbative contributions.
The calculations of the perturbative contribution has been extended to the next-to-next-to leading order (NNLO)  (see Ref. \refcite{Buras:2011we} and Refs. within)  greatly reducing the theoretical uncertainty, in particular
the matching scale uncertainty at NLO.
In the case of inclusive  $B \ra X_d \,  l^+ l^-$, the short distance  analysis is similar,
once one keeps the CKM suppressed terms in the operator expansion.
The  $  b \ra d l^+ l^-$ transition has been investigated  in the channel
 $ B^+ \ra \pi^+ \mu^+ \mu^-$  by the LHCb Collaboration,
observed for the first time in 2012
 using 1.0 $\mathrm{fb}^{-1}$ of data \cite{LHCb:2012de}.
The measured branching fraction
$ {\cal{B}} ( B^+ \ra \pi^+ \mu^+ \mu^-) = (2.3 \pm 0.6_{\mathrm{stat}} \pm 0.1_{\mathrm{syst}}) \times 10^{-8}$,
 is  compatible with  the predicted SM branching ratio \cite{Wang:2007sp}.
The LHCb Collaboration has also set upper limits, of order $10^{9}$,  on the branching fraction of the  lepton number violating process $ {\cal{B}} ( B^- \ra \pi^+ \mu^- \mu^-)$,
  searching for Majorana neutrinos with 3 $\mathrm{fb}^{-1}$ of data \cite{Aaij:2014aba}.

The inclusive $B \ra X_s l^+ l^-$  decays
 have been  measured at Belle \cite{Iwasaki:2005sy} and at BaBar \cite{Aubert:2004it}. Both results
find  branching ratios  of  order $10^{-6}$.
Using a sum over  exclusive modes as
the basis for extrapolation to the fully inclusive rate, a lepton-flavor-averaged inclusive
branching fraction has been recently measured by BaBar \cite{Lees:2013nxa}.
Also the lepton forward-backward asymmetry (assuming that it does not depend on  the lepton flavor) has been  recently measured, for the first time,    by Belle \cite{Sato:2014pjr} in inclusive
 $B \rightarrow X_s l^+ l^-$. The lepton forward-backward asymmetry had already  been
 measured in exclusive $ B \ra K^{\ast} l^+ l^-$ channels \cite{Wei:2009zv,Aubert:2008bi,Aaltonen:2011ja,Aaij:2011aa}.

The NLO and NNLO QCD corrections for inclusive decays can of course be also
used for the corresponding exclusive decays.
The kinematic available phase space in $ B \ra K^{(\ast)} \mu^+ \mu^-$
 is fully covered experimentally, with the
exception of the $ J/\psi$ 	 and $\psi^\prime$ resonances, which are removed
by cuts.
In the  exclusive channel $ B \ra K^{\ast} l^+ l^-$,
 a systematic theoretical description using QCD factorization in the heavy quark limit
is possible for small invariant di-lepton masses $q^2$, reducing the number of
independent form factors from 7 to 2  and allowing to calculate spectator effects \cite{Beneke:2001at}.
In the region of low $q^2$,  where the energy of the emitted meson is large in the
$B$ meson rest frame,   both LCSR
and soft collinear effective theory analyses have been performed, with some discrepancy in a form factor ratio\cite{Bell:2010mg}.
In
the high $q^2$ region,
 QCD factorization is
less justified, becoming invalid close to the endpoint of the spectrum at $q^2 = (m_B-m_K)^2$.
Alternative approaches have been developed, based on expansions whose scale is set by the large value of $q^2$ \cite{Buchalla:1998mt, Grinstein:2004vb, Beylich:2011aq, Bobeth:2010wg}.
The
 large $q^2$ region is the domain of election for
lattice QCD and
unquenched calculations of form factors  have been  performed, see e.g. \cite{Liu:2011raa, Zhou:2011be}. The more recent ones are   for $B \to K l l $ decays by HPQCD  \cite{Bouchard:2013eph, Bouchard:2013mia},    for $B_s \to K l \nu$ and $B \to K l l $ decays  by  Fermilab/MILC collaboration \cite{Liu:2013sya} and for $B_s \to \phi l^+ l^-$ and $B \to K^\star l^+ l^- $ decays by  the Cambridge  collaboration \cite{Horgan:2013hoa, Horgan:2013pva}.
In the same large $q^2$  region, ratios of  $ B \ra K^{\ast}$ form factors have been extracted from angular variables recently measured \cite{Aaltonen:2011ja, Ritchie:2013mx, Albrecht:2012is},  precisely the fraction
of longitudinally polarized vector mesons  and
the transverse asymmetry in the $ B \ra K^{\ast} l^+ l^-$ decay,  and found consistent  with  lattice results \cite{Hambrock:2012dg}.
In general,
the study of angular observables can be used advantageously in the $ B \ra K^{\ast} l^+ l^-$ decay, even to explore the possibility of new physics \cite{Becirevic:2012fy, Altmannshofer:2012az, Kosnik:2012dj, DescotesGenon:2012zf, Jager:2012uw, Matias:2014jua, Hiller:2013cza, Duraisamy:2013pia}.
The angular distribution
$ B \ra K^{\ast} (\ra K \pi)  l^+ l^-$ may be polluted by events coming from the distribution
$ B \ra K_0^{\ast} (\ra K \pi)  l^+ l^-$, where $K_0^\ast$ is a scalar meson resonance, and this possibility was  analyzed in Refs. \refcite{Becirevic:2012dp} \refcite{Matias:2012qz}.
An approach  within the LCSR has also been formulated to explore the S-wave generalized form factors
for the heavy meson transitions into the $\pi \pi$, $K \pi$
final states \cite{Meissner:2013hya}.

The measured branching fractions of the $ B^+ \ra K^{\ast+}   \mu^+ \mu^-$  and
$ B^0 \ra K^{0}   \mu^+ \mu^-$   decays \cite{Aaij:2014pli}
 all favour  lower values than
the SM expectations.
 LHCb  has recently reported the most precise measurement of the branching ratio for the
$ B^+ \ra K^+  \mu^+ \mu^-$ channel  to date, together with a study of its  angular distribution and differential branching fraction \cite{Aaij:2012vr}.
The differential decay rate  can be written as
\beq
\frac{1}{\Gamma} \frac{d \Gamma \left(  B^+ \ra K^+  \mu^+ \mu^-  \right) } {d \cos \theta} =
\frac{3}{4} \left( 1 -{\mathrm{F}_H}  \right) ( 1- \cos^2 \theta) +  \frac{1}{2} {\mathrm{F}_H}  + {\mathrm A_{FB}} \cos \theta
\eeq
 where  $\theta$ is  the angle between the $\mu^-$ and the $K^+$ in the rest frame of the
dimuon system. The two parameters, ${\mathrm{F}_H} $ and  the forward-backward asymmetry of the dimuon system, ${\mathrm A_{FB}} $, depend on  $q^2$. In the SM, ${\mathrm A_{FB}}$ is zero and ${\mathrm{F}_H} $ highly suppressed, and their  measured values are consistent with the SM expectations \cite{Aaij:2012vr, Aaij:2014tfa}. The consistency has been verified also for $ B^0 \ra K^0_s  \mu^+ \mu^-$ decay  \cite{Aaij:2014tfa}.
The differential branching fraction of the
$ B^+ \ra K^+  \mu^+ \mu^-$ decay is, however, consistently below the SM prediction at low $q^2$  \cite{Aaij:2012vr}.
The measurement of the CP asymmetry for the $ B^+ \ra K^+  \mu^+ \mu^-$ decay, instead, is consistent with the SM predictions \cite{Aaij:2013dgw}.
A broad peaking structure is observed in the dimuon spectrum of  the same decay,
 in
the kinematic region where the kaon has a low recoil against the dimuon system; the contribution of the resonant decay and of the interference of the yield for dimuon masses seems larger than theoretical estimates \cite{Aaij:2013pta}.
LHCb reports also the actual more precise determinations of  ${\mathrm A_{FB}} $ for the decay $B^0 \ra K^{\ast 0} \mu^+ \mu^-$   \cite{Aaij:2011aa}.
More recently, LHCb has announced a 3.7$\sigma$
local discrepancy in one of the $q^2$ bins
for one of the angular observables \cite{Aaij:2013qta}. This analysis has prompted a large number of theoretical investigations, searching for NP in  several frameworks, and  I will just  mention a few examples,
the Randall Sundrum approach \cite{Biancofiore:2014wpa}, $Z^\prime$ new coupling \cite{Altmannshofer:2014cfa, Altmannshofer:2013foa, Buras:2012jb, Buras:2013qja}, MSSM \cite{Behring:2012mv, Mahmoudi:2014mja}, Minimal Flavour Violation \cite{Hurth:2013ssa}, models based on the gauge group $SU(3)_c \times SU(3)_L \times U(1)_X$ \cite{Buras:2013dea}, NP contributions to Wilson coefficients  \cite{Descotes-Genon:2013wba},  scalar interactions \cite{Datta:2013kja}, and so on.

The decay $ B^0_s \ra \phi  \mu^+ \mu^-$, involving a $b \to s$
quark transition,
constitutes a flavour changing neutral current  process, and it has been observed at CDF for the first time \cite{Aaltonen:2011cn}. Recently, LHCb has performed an angular analysis and reported
the most precise determination of the
branching fraction to date.  The
branching fraction is slightly (about 1/2  on average) smaller than  the SM theory predictions \cite{Aaij:2013aln}.

The leptonic $B_s \to \mu^+ \mu^-$ is a  flavour changing neutral current  decay that  has  been recently measured for the first time \cite{Aaij:2012nna}.
These results have been improved and superseded  by the same LHCb Collaboration \cite{Aaij:2013aka}. The CMS
collaboration has also published a measurement of the  $ B_s \ra   \mu^+ \mu^-$ branching fraction \cite{Chatrchyan:2013bka}.
An average of the results from LHCb and CMS
has been performed \cite{CMSandLHCbCollaborations:2013pla} and the resulting branching
fraction is
$ {\cal{B}} (B_s \to \mu^+ \mu^-) = (2.9 \pm 0.7) \times 10^{-9}$, that agrees with the SM value,  updated not long ago\cite{Bobeth:2013uxa, DeBruyn:2012wk}.
However,
 deviations from the SM are still possible within
 the large experimental uncertainties.
 The $B_s \to \mu^+ \mu^-$ decay gives strong constraints on new physics, in particular on two Higgs doublet and Susy models, already at the current
experimental precision.

\section{Conclusions}

The experimental progress in semi-leptonic decays in the last years is impressive and the theoretical situation is rich in perspective. The perturbative calculations, in general, have reached a phase of maturity, and the larger theoretical errors are due to non-perturbative approaches.

Still awaiting firmly established solutions are  a few dissonances within the SM, such as the so-called ``1/2 vs 3/2 puzzle"  and ``gap problem", the possibility of flavour violation in decays into tauons,
the long standing tension between the inclusive and exclusive determination of $|V_{cb}|$  and  $|V_{ub}|$, and the recent $B \to K^\ast \mu^+ \mu^-$ anomaly.

The experimental analysis performed with the full Babar data sample for semileptonic $ B \to D^{(\ast)} \tau  \nu_\tau$
seems to confirm the excess with respect to the SM. Several NP scenarios have been devised, while on the experimental side it would be interesting to compare with  results from the full  Belle data sample.

Inclusive and exclusive values of $|V_{cb}|$ of comparable precision are available, although, on the whole, inclusive decays remain more precise.
Their tension  seems strengthened from the most recent
lattice calculation and confirmation  from different lattice groups would be welcome.
The discrepancy between the values of $|V_{ub}|$ is larger,  and the uncertainty on this value propagate on other $|V_{ub}|$ dependent observables (see e.g. Ref. \refcite{Buras:2012wv}). The error on the inclusive determination, around 4\%, is about one half than  the one on the exclusive determinations. Lattice computations that are now are in progress may help in the near future to reduce the exclusive uncertainties.
By increasing statistics, other channels, as $B \to \rho/\omega \,  l \nu$, but also baryonic ones, can become an interesting alternative to the traditional $B \to \pi l \nu$ decay for the exclusive extraction of $|V_{ub}|$.

Recently, the most active field has   been the one related to rare decays, prompted by the latest measurements of the angular correlations of the decay products in $B \to K^\ast \mu^+ \mu^-$, that display several
deviations from the SM predictions, the largest being around 3.7$\sigma$. Rare decays have always been
the search ground of election for NP in flavour physics, and further data and analyses
are longingly awaited  to ascertain if this channel shows real departures from SM expectations   or they
 result from statistical fluctuations and/or underestimated error uncertainties.

\section*{Acknowledgments}

The author  acknowledges partial support   by
Italian MIUR under project 2010YJ2NYW and INFN
under specific initiative QNP.

\bibliographystyle{ws-mpla}
\bibliography{VxbRef}

\end{document}